\documentclass{caosp309}

\usepackage{graphicx}

\usepackage{natbib}
\articleNo{301}
\pubyear{2020}
\volume{50}
\volnumber{3}
\firstpage{1}
\received{May 1, 2020}
\accepted{July 28, 2020}

\def\BibTeX{{\rm B\kern-.05em{\sc i\kern-.025em b}\kern-.08em
             T\kern-.1667em\lower.7ex\hbox{E}\kern-.125emX}}

\newcommand{\Msun}{M$_{\odot}$} 

\newcommand{\MESA}{\texttt{MESA}}

\begin{document}

%
\htitle{Pulsational characteristics of mass accreting stars}
\hauthor{A. Miszuda}

\title{Pulsational characteristics of mass accreting stars in close binary systems}


%
%
\author{
        A.\,Miszuda\orcid{0000-0002-9382-2542}
       }

%
\institute{
         Nicolaus Copernicus Astronomical Centre, Polish Academy of Sciences \\ 
         Bartycka 18, PL-00-716 Warsaw, Poland \\ 
         \email{amiszuda@camk.edu.pl}
          }

\date{March 8, 2003}

\maketitle

\begin{abstract}
Eclipsing binaries with pulsating components are a distinct subclass of binaries, merging orbital and pulsational analyses. In recent years that subclass led to the definition of a newly formed branch of tidal asteroseismology. 
While single-star pulsators are well understood, the effects of binarity and possible mass transfer on pulsational characteristics, particularly in mass-gaining stars, remain to be systematically explored. Here, I present preliminary results on the asteroseismic properties of a mass-accreting model for a 10\,\Msun\ $\beta$\,Cephei-type star.

\end{abstract}

%

\section{Introduction}
\label{introduction}

The recent years have brought an enormous progress in development and understanding of tidal asteroseismology. \cite{Handler2020} have reported the first ever observed case of a single p-mode tidally trapped $\delta$\,Sct pulsation in a close binary star system, HD 74423.
That discovery was followed by \cite{Kurtz2020}, who discovered at least four separate tidally trapped oscillation modes in CO\,Cam system. As shown by \cite{Fuller2020}, that exotic behaviour has its roots in mode coupling.


Further advancements have been made by recent studies, including \cite{Miszuda2021, Miszuda2022}, who explored the role of mass transfer in affecting radial modes and mode instabilities in $\delta$\,Sct stars. Expanding on this, \cite{Wagg2024} used Slowly Pulsating B (SPB) star models to show that mass-accreting binaries often exhibit amplitude variations in their period spacing patterns, with some patterns being out of phase. Additionally, \cite{Rui2021}, \cite{Bellinger2024}, and \cite{Henneco2024} demonstrated that asteroseismology can be used to distinguish merger products, by investigating how mass transfer affects the pulsation spectra of merger remnants. These studies collectively highlight the potential of asteroseismology in unveiling the complex effects of mass transfer and stellar mergers on the pulsational properties of stars.

The motivation for this research lies in the fact that stars within the mass range of $\beta$\,Cephei-type stars (8 -- 18\,M$_{\odot}$) can be frequently found in binary or multiple systems \citep{Duchene2013}, where mass transfer can significantly alter their evolution. This provides excellent conditions for studying the interplay between stellar pulsations and mass transfer.


\section{Methodology}
\label{methodology}
Using \MESA\ \citep[Modules for Experiments in Stellar Astrophysics,][version 23.05.1]{Paxton2011}, I built a toy model of a binary system with $M_{\rm don,0} = 10$\,\Msun\ and $M_{\rm acc,0} = 7$\,\Msun\ initial masses in a $P_{\rm orb, 0} = 3$\,d circular orbit. The binary was evolved through a phase of mass transfer until mass ratio reversal, i.e., $M_{\rm don} = 7$\,\Msun\ and $M_{\rm acc} = 10$\,\Msun. At this point, all binary interactions were halted, and the mass-accreting star evolved as a single star until central hydrogen exhaustion.

Solar metallicity was assumed for each component, with exponential overshooting mixing applied to the hydrogen-burning core, with $f_{\rm ov} = 0.02$, following the \cite{Herwig2000} prescription. 
Convective, semi-convective, and thermohaline mixing were incorporated with coefficients $\alpha_{\rm MLT} = 0.5$, $\alpha_{\rm SC} = 0.1$, and $\alpha_{\rm TH} = 1$, respectively.

For the sake of this paper I selected a model at $X_c = 0.1$ for the mass-accreting star and calculated theoretical pulsations for $\ell=0,1,2$ modes using \texttt{GYRE} \citep{Townsend2018}. This model is approximately 39.31\,Myr old and is well past the mass transfer phase, which started at $X_c = 0.52$ (23.64\,Myr) and ended at $X_c = 0.59$ (23.65\,Myr). All computations were repeated for a single $M = 10$\,\Msun\ star as a basis for comparison.

\section{Results}
\label{results}

\begin{figure}
\centerline{\includegraphics[width=\textwidth,clip=]{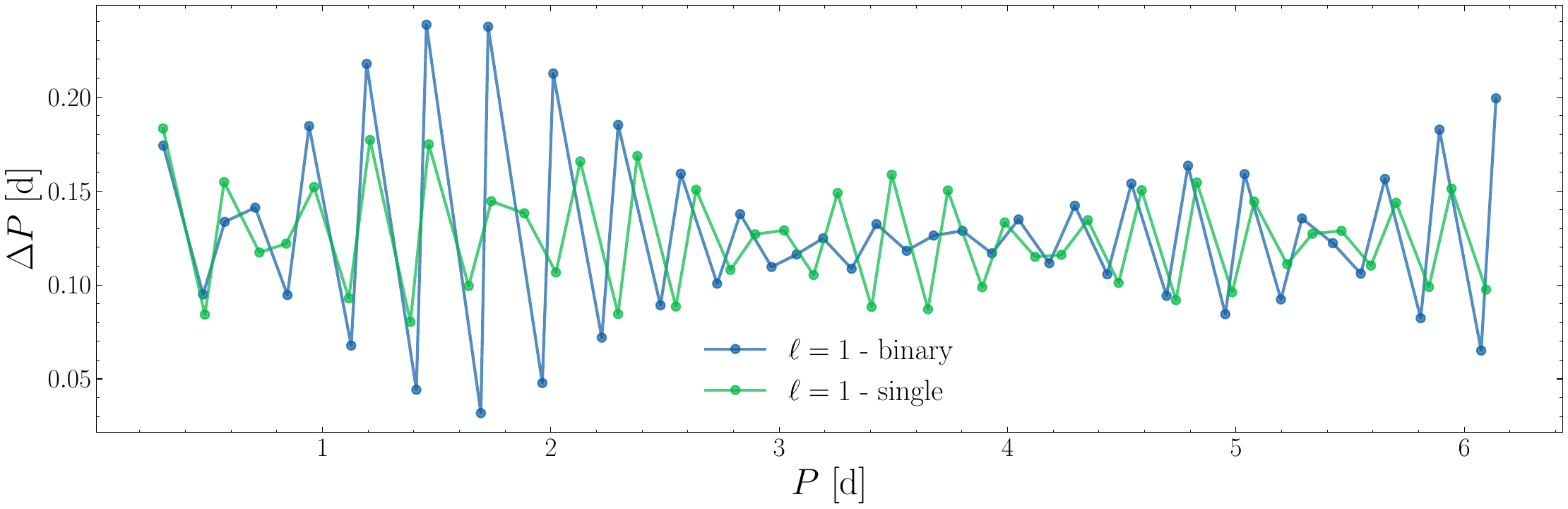}}
\caption{The period spacing patterns of the $\ell = 1$ g-modes for $M = 10$\,\Msun, $X_c = 0.1$ single (green line) and binary (blue line) models.}
\label{dP}
\end{figure}

For each pulsational model, I calculated the period spacing, $\Delta P$, using $\ell = 1$ g-modes frequencies. As shown in Fig.\,\ref{dP}, the main differences between the single-star and mass-accreting models are in amplitude deviation and phase shifts in the period spacing patterns. The single-star model exhibits a lower amplitude compared to the binary case. Additionally, the mass-accreting model is out of phase for periods between 3-4 days. A similar behaviour was observed by \cite{Wagg2024} for SPB star models.

Next, the $\nu_{s}/\nu_{b}$ frequency ratios were computed for $\ell = 0,1,2$ p-modes, where $\nu_s$ is the frequency from the single-star model and $\nu_b$ is the frequency from the binary model, ensuring that they correspond to the same radial orders $n_{\rm pg}$ in both cases. The most noticeable deviations appear at low radial orders, between $n_{\rm pg} = 0$ and $n_{\rm pg} = 10$, suggesting that low-order p-modes may be sensitive to differences in the internal structure of stars that have undergone mass transfer. This behaviour will be the subject of in-depth study in our future work.

\begin{figure}
\centerline{\includegraphics[width=\textwidth,clip=]{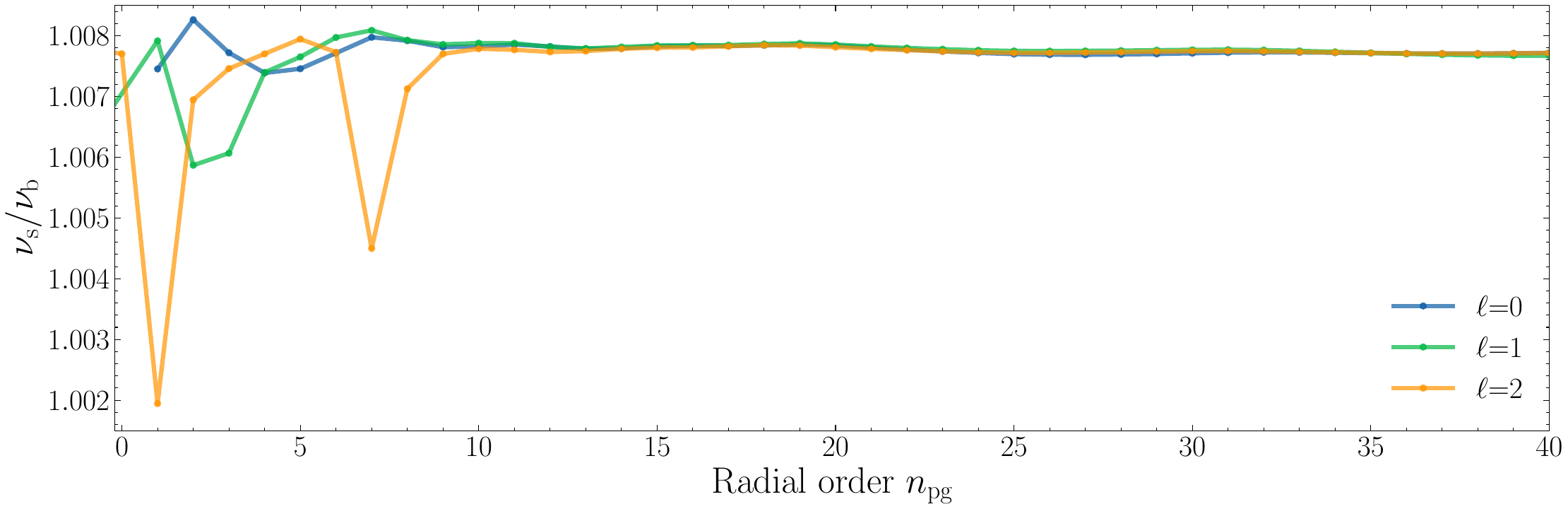}}
\caption{The frequency ratio patterns of the $\ell = 0,1,2$ p-modes for $M = 10$\,\Msun, $X_c = 0.1$ models.}
\label{df}
\end{figure}

\section{Summary}
\label{summary}
The results presented in this proceedings suggest that mass accretion leads to measurable deviations in the pulsational properties of stars. This ongoing work will be expanded in a forthcoming paper by Miszuda \& Guo (in preparation).

\acknowledgements
\noindent This work was supported by the Polish National Science Centre (NCN), grant number 2021/43/B/ST9/02972.

\bibliography{biblio}

\begin{thebibliography}{13}
\expandafter\ifx\csname natexlab\endcsname\relax\def\natexlab#1{#1}\fi

\bibitem[{{Bellinger} {et~al.}(2024){Bellinger}, {de Mink}, {van Rossem}, \& {Justham}}]{Bellinger2024}
{Bellinger}, E.~P., {de Mink}, S.~E., {van Rossem}, W.~E., \& {Justham}, S., {The Potential of Asteroseismology to Resolve the Blue Supergiant Problem}. 2024, {\it \apjl}, {\bf 967}, L39, DOI: 10.3847/2041-8213/ad4990

\bibitem[{{Duch{\^e}ne} \& {Kraus}(2013)}]{Duchene2013}
{Duch{\^e}ne}, G. \& {Kraus}, A., {Stellar Multiplicity}. 2013, {\it \araa}, {\bf 51}, 269, DOI: 10.1146/annurev-astro-081710-102602

\bibitem[{{Fuller} {et~al.}(2020){Fuller}, {Kurtz}, {Handler}, \& {Rappaport}}]{Fuller2020}
{Fuller}, J., {Kurtz}, D.~W., {Handler}, G., \& {Rappaport}, S., {Tidally trapped pulsations in binary stars}. 2020, {\it \mnras}, {\bf 498}, 5730, DOI: 10.1093/mnras/staa2376

\bibitem[{{Handler} {et~al.}(2020){Handler}, {Kurtz}, {Rappaport}, {Saio}, {Fuller}, {Jones}, {Guo}, {Chowdhury}, {Sowicka}, {Kahraman Ali{\c{c}}avu{\c{s}}}, {Streamer}, {Murphy}, {Gagliano}, {Jacobs}, \& {Vanderburg}}]{Handler2020}
{Handler}, G., {Kurtz}, D.~W., {Rappaport}, S.~A., {et~al.}, {Tidally trapped pulsations in a close binary star system discovered by TESS}. 2020, {\it Nature Astronomy}, {\bf 4}, 684, DOI: 10.1038/s41550-020-1035-1

\bibitem[{{Henneco} {et~al.}(2024){Henneco}, {Schneider}, {Hekker}, \& {Aerts}}]{Henneco2024}
{Henneco}, J., {Schneider}, F.~R.~N., {Hekker}, S., \& {Aerts}, C., {Merger seismology: Distinguishing massive merger products from genuine single stars using asteroseismology}. 2024, {\it \aap}, {\bf 690}, A65, DOI: 10.1051/0004-6361/202450508

\bibitem[{{Herwig}(2000)}]{Herwig2000}
{Herwig}, F., {The evolution of AGB stars with convective overshoot}. 2000, {\it \aap}, {\bf 360}, 952

\bibitem[{{Kurtz} {et~al.}(2020){Kurtz}, {Handler}, {Rappaport}, {Saio}, {Fuller}, {Jacobs}, {Schmitt}, {Jones}, {Vanderburg}, {LaCourse}, {Nelson}, {Kahraman Ali{\c{c}}avu{\c{s}}}, \& {Giarrusso}}]{Kurtz2020}
{Kurtz}, D.~W., {Handler}, G., {Rappaport}, S.~A., {et~al.}, {The single-sided pulsator CO Camelopardalis}. 2020, {\it \mnras}, {\bf 494}, 5118, DOI: 10.1093/mnras/staa989

\bibitem[{{Miszuda} {et~al.}(2022){Miszuda}, {Ko{\l}aczek-Szyma{\'n}ski}, {Szewczuk}, \& {Daszy{\'n}ska-Daszkiewicz}}]{Miszuda2022}
{Miszuda}, A., {Ko{\l}aczek-Szyma{\'n}ski}, P.~A., {Szewczuk}, W., \& {Daszy{\'n}ska-Daszkiewicz}, J., {The eclipsing binary systems with {\ensuremath{\delta}} Scuti component - II. AB Cas}. 2022, {\it \mnras}, {\bf 514}, 622, DOI: 10.1093/mnras/stac1197

\bibitem[{{Miszuda} {et~al.}(2021){Miszuda}, {Szewczuk}, \& {Daszy{\'n}ska-Daszkiewicz}}]{Miszuda2021}
{Miszuda}, A., {Szewczuk}, W., \& {Daszy{\'n}ska-Daszkiewicz}, J., {The eclipsing binary systems with {\ensuremath{\delta}} Scuti component - I. KIC 10661783}. 2021, {\it \mnras}, {\bf 505}, 3206, DOI: 10.1093/mnras/stab1597

\bibitem[{{Paxton} {et~al.}(2011){Paxton}, {Bildsten}, {Dotter}, {Herwig}, {Lesaffre}, \& {Timmes}}]{Paxton2011}
{Paxton}, B., {Bildsten}, L., {Dotter}, A., {et~al.}, {Modules for Experiments in Stellar Astrophysics (MESA)}. 2011, {\it \apjs}, {\bf 192}, 3, DOI: 10.1088/0067-0049/192/1/3

\bibitem[{{Rui} \& {Fuller}(2021)}]{Rui2021}
{Rui}, N.~Z. \& {Fuller}, J., {Asteroseismic fingerprints of stellar mergers}. 2021, {\it \mnras}, {\bf 508}, 1618, DOI: 10.1093/mnras/stab2528

\bibitem[{{Townsend} {et~al.}(2018){Townsend}, {Goldstein}, \& {Zweibel}}]{Townsend2018}
{Townsend}, R.~H.~D., {Goldstein}, J., \& {Zweibel}, E.~G., {Angular momentum transport by heat-driven g-modes in slowly pulsating B stars}. 2018, {\it \mnras}, {\bf 475}, 879, DOI: 10.1093/mnras/stx3142

\bibitem[{{Wagg} {et~al.}(2024){Wagg}, {Johnston}, {Bellinger}, {Renzo}, {Townsend}, \& {de Mink}}]{Wagg2024}
{Wagg}, T., {Johnston}, C., {Bellinger}, E.~P., {et~al.}, {The asteroseismic imprints of mass transfer. A case study of a binary mass-gainer in the SPB instability strip}. 2024, {\it \aap}, {\bf 687}, A222, DOI: 10.1051/0004-6361/202449912

\end{thebibliography}

\end{document}